\documentclass[prl,12pt]{revtex4}

\usepackage[english]{babel}
\usepackage[intlimits]{amsmath}
\usepackage[dvips]{graphicx}
\begin{document} 

\title[Burnt-Bridge Model]{Dynamics of Molecular Motors in Reversible Burnt-Bridge Models}
\author{Maxim N. Artyomov\ddag, Alexander Yu. Morozov\ddag\dag, and Anatoly B. Kolomeisky\dag}
\affiliation{\dag Department of Chemistry, Rice University, Houston, TX 77005, USA}
\affiliation{\ddag Department of Chemistry, Massachusetts Institute of Technology, Cambridge, MA 02139, USA}
\affiliation{\ddag\dag Department of Physics and Astronomy, University of
  California, Los Angeles, Los Angeles, CA 90095, USA}

\begin{abstract}
Dynamic properties of molecular motors whose motion is powered by interactions with specific lattice bonds are studied theoretically with the help of discrete-state stochastic ``burnt-bridge'' models. Molecular motors are depicted as random walkers that can destroy or rebuild periodically distributed weak connections (``bridges'') when crossing them, with probabilities $p_1$ and $p_2$ correspondingly. Dynamic properties, such as velocities and dispersions, are obtained in exact and explicit form for arbitrary values of parameters $p_1$ and $p_2$. For the unbiased random walker, reversible burning of the bridges results in a biased directed motion with a dynamic transition observed at very small concentrations of bridges. In the case of backward biased molecular motor its backward velocity is reduced and a reversal of the direction of motion is observed for some range of parameters. It is also found that the dispersion demonstrates a complex, non-monotonic behavior with large fluctuations for some set of parameters. Complex dynamics of the system is discussed by analyzing the behavior of the molecular motors near burned bridges.  
\end{abstract}

\maketitle

\section{Introduction}

In recent years an increased attention has been devoted to investigations of molecular motors, also known as motor proteins, that are crucial in many cellular processes \cite{AR}. They transform chemical energy into the mechanical motion in non-equilibrium conditions. For most of molecular motors their motion along linear molecular tracks is fueled by the hydrolysis of adenosine triphosphate (ATP) or related compounds. It was suggested that a different
mechanism is employed to power the motion of a protein collagenase along collagen fibrils \cite{saffarian04,saffarian06}. It probably utilizes the collagen proteolysis, cleaving the filament at specific sites. As the collagenase molecule is unable to cross the already broken bond, it leads to the biased diffusion along the filament. However, full understanding of mechanisms of collagenase motion is still not available.

It was proposed that a good description of the collagenase dynamics could be provided by the so-called ``burnt-bridge model'' (BBM) \cite{saffarian04,saffarian06,mai01,antal05,mp,ampk,mk,amk}.  In this model, the motor protein is depicted as a random walker that translocates along the one-dimensional lattice that consists of strong and weak bonds. While the strong bonds remain unaffected if crossed by the walker in any direction, the weak ones (termed ``bridges'') might be broken (or ``burnt'') with a probability $0<p_1\le 1$ when crossed in the specific direction, and the walker cannot cross the burnt bridges again, unless they are restored, which can occur with probability $0<p_2\le 1$. In
Refs. \cite{mp,ampk} an analytical approach was developed which permitted us to derive the explicit formulas for molecular motor velocity $V(c,p_1)$ and diffusion constant  $D(c,p_1)$ for the entire ranges of burning probability $0< p_1\le 1$ and concentration of the bridges $0<c\le 1$ which were also confirmed by extensive Monte Carlo computer simulations. This theoretical method has been applied to several problems with  periodic bridge distribution. However, the results in \cite{mp,ampk} have been obtained only for irreversible bridge burning (bridge recovery probability was taken to be $p_2=0$), and also for unbiased random walker between bridges (equal forward and backward transition rates). In present work, we generalize our approach to allow for the possibility of bridge recovery as well as unequal hopping rates on the sites between bridges. It is more realistic to consider systems with reversible action of motor proteins since they are catalysts that equally accelerate both forward and backward biochemical transitions \cite{AR}.

\section{Model}

According to our model, we view a motor protein as a random walker moving along an infinite one-dimensional lattice with forward and backward transition rates being $u$ and $w$ correspondingly, as illustrated in Fig. 1. The lattice
spacing size is set to be equal to one. The lattice is composed of strong and weak bonds. There is no interaction between the random walker and strong bonds, however crossing the bridge in the forward direction (from left to right) leads to its burning with the probability $p_1$, while the particle moves with the rate $u$. After the weak link is destroyed, the walker is assumed to be on the right side of it. When the particle is trying to cross a broken bond, the bridge can be recovered with the probability $p_2$, while  the particle moves to the left with rate $w$. It is assumed that initially, at $t=0$, all bridges are intact.

The details of breaking  weak bonds in BBM have a strong effect on the dynamic properties of motor proteins \cite{mp}. There are two different possibilities of bridge burning. In the first variant (the so called ``forward BBM''), the weak bond is broken when crossed from left to right, but the intact bridge is not affected when the particle moves from right to left. Thus the bridge recovery may occur if the walker attempts to cross a burnt bridge from right to left. In the second variant (named ``forward-backward BBM''), the weak link is destroyed if crossed in either direction \cite{mai01,antal05}. Both variants are identical for $p_1=1$, however for $p_1<1$ the dynamics is different in two burning scenarios, as was shown in $p_2=0$ case \cite{mp}, although mechanisms are still the same. For reasons of simplicity, below we will only consider forward BBM, even though forward-backward BBM can also be solved using the same method. 

\begin{figure}[tbp]\label{Fig1}
\centering
\includegraphics[scale=0.6,clip=true]{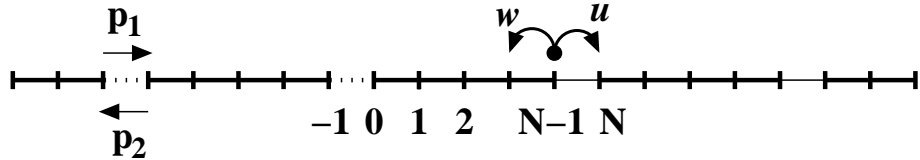}
\caption{A schematic picture of the motion of a molecular motor in the reversible burnt-bridge model. Thick solid lines depict strong links, while thin solid lines represent periodically distributed weak links (bridges). Dotted lines are for already burnt bridges. }
\end{figure}

There are five parameters that specify the dynamics of  molecular motors in BBM: the probabilities $p_1$, $p_2$, the concentration of bridges $c$, as well as transition rates $u$ and $w$. The dynamic properties of the walker are also strongly influenced by the distribution of weak bonds \cite{antal05}. Below we will study the case of periodically distributed bridges, when their concentration is $c=1/N$ and the weak bonds are located between the lattice sites with the coordinates $kN-1$ and $kN$, with integer $k$ (see Fig. 1). This description is  more realistic for collagenases' dynamics \cite{saffarian04,saffarian06}. The model below will be studied using continuous time analysis as it better describes chemical transitions in motor proteins \cite{mp}.

\section{Dynamic properties for BBM with bridge recovery}

\subsection{Velocity}

To find the walker's velocity, we generalize the method used in \cite{mp} (for irreversible bridge burning, i.e. $p_2=0$) to allow for non-zero probability $p_2$. We introduce a probability $R_j(t)$ that the random walker is found $j$ sites apart from the last burnt bridge at time $t$. The probabilities $R_j(t)$ arise if the system is viewed in moving coordinate frame with the last burnt bridge always at the origin, as illustrated by a reduced chemical  kinetic scheme shown in Fig. 2.

\begin{figure}[tbp]\label{Fig2}
\centering
\includegraphics[scale=0.6,clip=true]{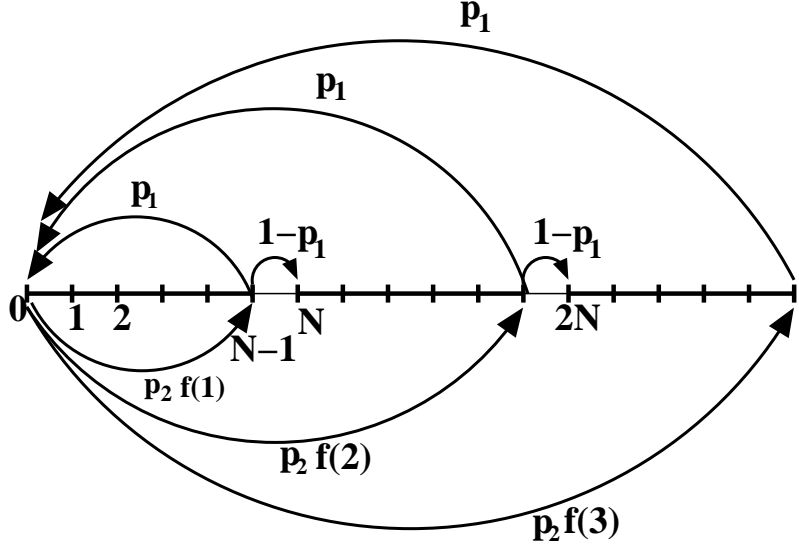}
\caption{Reduced kinetic scheme for continuous-time forward BBM with bridge recovery (only transition rates not equal to $u$, $w$ are shown). The origin is the right end of the last burnt bridge. Parameters are described in detail in the text.  }
\end{figure}

The dynamics of the system is determined by a set of master equations: 
\begin{equation}\label{me1}
\frac{d R_{kN+i}(t)}{dt} = uR_{kN+i-1}(t)+wR_{kN+i+1}(t)-(u+w)R_{kN+i}(t),
\end{equation}
for $k=0,1,2,\cdots$ and $i=1,2,\cdots,N-2$; and
\begin{equation}\label{me2}
\frac{d R_{kN+N-1}(t)}{dt} = uR_{kN+N-2}(t)+wp_{2}f(k+1)R_0(t)+wR_{(k+1)N}(t)-(u+w)R_{kN+N-1}(t),
\end{equation}
\begin{equation}\label{me3}
\frac{d R_{(k+1)N}(t)}{dt}=(1-p_1)uR_{kN+N-1}(t)+wR_{(k+1)N+1}(t)-(u+w)R_{(k+1)N}(t),
\end{equation}
with $k=0,1,2,\cdots$ for both Eqs. (\ref{me2}) and (\ref{me3}). Also  at the origin we have
\begin{equation}\label{me4}
\frac{d R_{0}(t)}{d t}=p_1 u \sum_{k=1}^{\infty} [R_{kN-1}(t)] +wR_{1}(t)-uR_{0}(t)-wp_2 R_{0}(t).
\end{equation}
In Eq. (\ref{me2}) we introduced a function $f(k)$ as a probability that next to the last burnt bridge is $k$ periods to the left from the last burnt bridge. It satisfies the condition $\sum \limits_{k=1}^{\infty}f(k)=1$, which is reflected in Eq. (\ref{me4}). The system of equations (\ref{me1})-(\ref{me4}) is to be solved in the stationary-state limit (at large times)  when $d R_j(t)/dt=0$ is satisfied, and we denote $R_j(t\rightarrow\infty)\equiv R_j$ in what follows. By definition, it can be argued that 
\begin{equation}\label{f(k)1}
f(k)=\frac{p_1 R_{kN-1}}{\sum \limits_{k=1}^{\infty}p_1 R_{kN-1}}.
\end{equation}
Solving the system (\ref{me1})-(\ref{me4}) can be facilitated by rewriting Eq. (\ref{f(k)1}) in a more convenient form. To this end, we note that based on the results from Refs. \cite{mp,mk} it is reasonable to assume that  $R_{kN+i}$ is of the form 
\begin{equation}\label{ryw}
R_{kN+i}=y^k W(i),
\end{equation}
where $y$ and $W$ are some functions of $p_1$, $p_2$, $u$, $w$ and $N$. Furthermore, $y$ is $i$ and $k$-independent, while $W$ depends on $i$ (but not on $k$). The advantage of the ansatz (\ref{ryw}) is that it leads to a simpler form of $f(k)$:
\begin{equation}\label{f(k)2}
f(k)=y^{k-1}(1-y),
\end{equation}
as follows from Eq. (\ref{f(k)1}). We proceed to solve Eqs. (\ref{me1})-(\ref{me4}) with $f(k)$ in Eq. (\ref{me2}) given by the expression (\ref{f(k)2}).

Introducing a parameter $\beta\equiv u/w$, it can be easily verified that 
\begin{equation}\label{c1c2}
R_{kN+i}=C_1(k)+C_2(k)\beta^i
\end{equation}
with arbitrary $C_1(k)$, $C_2(k)$ solves Eq. (\ref{me1}). Utilizing  Eq. (\ref{c1c2}), $C_1(k)$, $C_2(k)$ can be expressed in terms of functions $R_{kN}$ and $R_{kN+1}$ (first two points of the period). Then Eq. (\ref{c1c2}) leads to \begin{equation}\label{rdelta}
R_{kN+i}=R_{kN}-\frac{w}{u-w}\left[1-\beta^i\right]\Delta_k,
\end{equation}
where we defined 
\begin{equation}\label{}
\Delta_k=R_{kN+1}-R_{kN}.
\end{equation}
In expression (\ref{rdelta}) it was assumed that $k=0,1,2,\cdots$ and $i=0,1,\cdots,N-1$. Parameters $R_{kN}$ and $\Delta_k$ are to be determined from Eqs. (\ref{me2}) and (\ref{me3}). Substituting Eq. (\ref{rdelta}) in (\ref{me2}) and (\ref{me3}) yields 
\begin{eqnarray}\label{eq1}
uR_{kN}-\frac{uw}{u-w}\left[1-\beta^{N-2}\right]\Delta_k + wp_2 y^k(1-y)R_0+wR_{(k+1)N}-(u+w)R_{kN} \nonumber \\ 
+(u+w)\frac{w}{u-w}\left[1-\beta^{N-1}\right]\Delta_k=0,
\end{eqnarray} 
\begin{eqnarray}\label{eq2}
(1-p_1)uR_{kN}-(1-p_1)\frac{uw}{u-w}\left[1-\beta^{N-1}\right]\Delta_k+wR_{(k+1)N} \nonumber \\ 
-\frac{w^2}{u-w}\left[1-\beta\right]\Delta_{k+1}-(u+w)R_{(k+1)N}=0.
\end{eqnarray} 
Eq. (\ref{eq1}) also yields 
\begin{equation}\label{delta1}
\Delta_k=\frac{u-w}{w(1-\beta^N )}[R_{kN}-R_{(k+1)N}-p_2y^k(1-y)R_0].
\end{equation}
Substituting (\ref{delta1}) into Eq. (\ref{eq2}) results in 
\begin{equation}\label{abc1}
aR_{kN}+bR_{(k+1)N}+cR_{(k+2)N}+(d+ey+fy^2)R_0y^k=0,
\end{equation}
where 
\begin{equation}\label{a}
a=(1-p_1)u\left[1-\frac{1-\beta^{N-1}}{1-\beta^N}\right],
\end{equation}
\begin{equation}\label{}
b=-u+(1-p_1)u\frac{1-\beta^{N-1}}{1-\beta^N}-w\frac{1-\beta}{1-\beta^N},
\end{equation}
\begin{equation}\label{eq_c}
c=w\frac{1-\beta}{1-\beta^N},
\end{equation}
\begin{equation}\label{}
d=p_2(1-p_1)u\frac{1-\beta^{N-1}}{1-\beta^N},
\end{equation}
\begin{equation}\label{}
e=-p_2(1-p_1)u\frac{1-\beta^{N-1}}{1-\beta^N}+wp_2\frac{1-\beta}{1-\beta^N},
\end{equation}
and
\begin{equation}\label{f}
f=-wp_2\frac{1-\beta}{1-\beta^N}.
\end{equation}
Using Eq. (\ref{ryw}), it follows that 
\begin{equation}\label{rkn}
R_{kN}=R_0 y^k,
\end{equation}
and Eq. (\ref{abc1}) turns into
\begin{equation}\label{abc2}
(c+f)y^2+(b+e)y+(a+d)=0.
\end{equation}
Therefore 
\begin{equation}\label{y}
y=\frac{-(b+e)-\sqrt{(b+e)^2-4(a+d)(c+f)}}{2(c+f)},
\end{equation}
where we selected the solution of (\ref{abc2}) such that $0\le y<1$. With the help of (\ref{rkn}) we obtain from Eq. (\ref{delta1})
\begin{equation}\label{delta2}
\Delta_k=\frac{u-w}{w\left\{1-\beta^N\right\}}R_0y^k(1-y)(1-p_2).
\end{equation}
Thus (\ref{rdelta}) yields
\begin{equation}\label{rkni}
R_{kN+i}=R_0 y^k\left[1-(1-p_2)(1-y)\frac{1-\beta^i}{1-\beta^N}\right].
\end{equation}
Eq. (\ref{rkni}) with $y$ given by (\ref{y}) solves the system of Eqs. (\ref{me1})-(\ref{me4}) in the stationary state limit. We note that although the Eq. (\ref{me4}) was not used to find this solution,  it was numerically verified that every equation in the system (\ref{me1})-(\ref{me4}) is indeed solved by expressions (\ref{rkni}) and (\ref{y}). 

Parameter $R_0$ needed to find the velocity is found from Eq. (\ref{rkni}) combined with the normalization condition $\sum \limits_{k=0}^{\infty}\sum \limits_{i=0}^{N-1}R_{kN+i}=1$, which produces 
\begin{equation}\label{r0}
R_0=(1-y)\left\{N-(1-p_2)(1-y)\left[\frac{N}{1-\beta^N}-\frac{1}{1-\beta}\right]\right\}^{-1}.
\end{equation}
The mean velocity of the walker is given by \cite{mp} 
\begin{equation}\label{vel}
V=\sum \limits_{j=0}^{\infty}(u_j-w_j)R_j=(u-wp_2)R_0+\left[\sum \limits_{k=0}^{\infty}\sum \limits_{i=0}^{N-1}(u-w)R_{kN+i}\right]-(u-w)R_0,
\end{equation}
which results in a simple relation,
\begin{equation}\label{vel1}
V=w(1-p_2)R_0+(u-w).
\end{equation}
In Eq. (\ref{vel1}), $R_0$ is given by (\ref{r0}) with $y$ from the expression (\ref{y}). 

It can be shown that in the limit of $u\rightarrow 1$, $w\rightarrow 1$, and $p_2\rightarrow 0$ Eq. (\ref{vel1}) reproduces the result obtained earlier in the Ref.  \cite{mp} for the BBM with $u=w=1$ and $p_2=0$. Also, Eq. (\ref{vel1}) simplifies considerably in the limiting case of $p_1=1$ (deterministic bridge burning) when $y=0$. In the case of $p_1=1$ we obtained $V(u,w,p_2,N)$ in \cite{amk} using the Derrida's method \cite{derrida83} and our general result given in Eq. (\ref{vel1}) agrees with it in the $p_1\rightarrow 1$ limit, as was numerically verified.

\subsection{Diffusion Coefficient}

The diffusion coefficient is found by generalizing the method developed in \cite{ampk} (where we found dynamic properties of the random walker in BBM with $u=w=1$, $p_2=0$ and periodic bridge distribution), allowing for $0<p_2\le 1$ and $u\neq w$. We define $P_{kN+i,m}(t)$ as the probability that at time $t$ the random walker is located at point $x=kN+i$ ($i=0,1,\cdots,N-1$), the right end of the last burnt bridge being at the point $mN$. Parameters $m$ and $k \ge 0$ assume integer values.

The dynamics of the system is described by a set of Master equations:
\begin{equation}\label{p1}
\frac{d P_{mN,m}(t)}{d t}=wP_{mN+1,m}(t)-(u+p_2w)P_{mN,m}(t)+p_1u \sum_{m^{\prime}=-\infty}^{m-1} P_{mN-1,m^{\prime}}(t),
\end{equation}
for $k=i=0$,
\begin{eqnarray}\label{p2}
\frac{d P_{mN+kN+(N-1),m}(t)}{dt}=wP_{mN+(k+1)N,m}(t)+uP_{mN+kN+(N-2),m}(t) \nonumber \\ 
+p_2wf(k+1)P_{mN+(k+1)N,m+k+1}(t)-(u+w)P_{mN+kN+(N-1),m}(t),
\end{eqnarray}
for $k\ge 0$ and $i=N-1$ [with $f(k+1)$ the same as in (\ref{me2})],
\begin{equation}\label{p3}
\frac{d P_{mN+kN,m}(t)}{dt}=wP_{mN+kN+1,m}(t)+(1-p_1)uP_{mN+kN-1,m}(t)-(u+w)P_{mN+kN,m}(t),
\end{equation}
for $k\ge 1$ and $i=0$, and
\begin{equation}\label{p4}
\frac{d P_{mN+kN+i,m}(t)}{dt}=wP_{mN+kN+i+1,m}(t)+uP_{mN+kN+i-1,m}(t)-(u+w)P_{mN+kN+i,m}(t)
\end{equation}
for $k \ge 0$ and $i=1,\cdots,N-2$.

We observe that
\begin{equation}\label{r00}
R_{kN+i}(t)=\sum \limits_{m=-\infty}^{+\infty}P_{(m+k)N+i,m}(t), \quad  (k \ge 0; \ i=0,1,\cdots,N-1),
\end{equation}
where $R_{kN+i}(t)$ is the probability for the random walker to be found $kN+i$ sites apart from the last burnt bridge at time $t$, which was used above to find the walker's velocity and is given by Eq. (\ref{rkni}). Plugging (\ref{r00}) into Eqs. (\ref{p1}) - (\ref{p4}) results in the equations (\ref{me1}) - (\ref{me4}) for $R_{kN+i}$, thus obtained by a different method.

In accordance with \cite{ampk,derrida83},  we introduce auxiliary functions $S_{kN+i}(t)$,
\begin{equation}\label{s0}
S_{kN+i}(t)=\sum\limits_{m=-\infty}^{+\infty}(mN+kN+i)P_{mN+kN+i,m}(t), \quad (k \ge 0 \ i=0,1,\cdots,N-1).
\end{equation}
The system of equations describing the time evolution of the functions $S_{kN+i}(t)$ results from Eqs. (\ref{s0})  and Eqs. (\ref{p1}) - (\ref{p4}). It was obtained that
\begin{equation}\label{s1}
\frac{d S_{0}(t)}{d t}=wS_1(t)-(u+p_2w)S_0(t)+p_1u\sum\limits_{\alpha=1}^{\infty}S_{\alpha N-1}(t)+p_1u\sum\limits_{\alpha=1}^{\infty}R_{\alpha N-1}(t)-wR_1(t),
\end{equation}
\begin{eqnarray}\label{s2}
\frac{d S_{kN+(N-1)}(t)}{d t}=wS_{(k+1)N}(t)+uS_{kN+(N-2)}(t)+p_2wf(k+1)[S_0-R_0] \nonumber \\ -(u+w)S_{kN+(N-1)}(t)-wR_{(k+1)N}(t)+uR_{kN+(N-2)}(t),
\end{eqnarray}
for $k\ge 0$,
\begin{equation}\label{s3}
\frac{d S_{kN}(t)}{d t}=wS_{kN+1}(t)+(1-p_1)uS_{kN-1}(t)-(u+w)S_{kN}(t)+(1-p_1)uR_{kN-1}(t)-wR_{kN+1}(t)
\end{equation}
for $k \ge 1$ and 
\begin{equation}\label{s4}
\frac{d S_{kN+i}(t)}{d t}=wS_{kN+i+1}(t)+uS_{kN+i-1}(t)-(u+w)S_{kN+i}(t)+uR_{kN+i-1}(t)-wR_{kN+i+1}(t)
\end{equation} 
for $k \ge 0$ and $i=1,\cdots,N-2$.

At $t\rightarrow\infty$ the solutions of Eqs. (\ref{s1}) - (\ref{s4}) are sought in the form
\begin{equation}\label{sj}
S_j(t)=a_j t+T_{j},
\end{equation} 
where $a_j$ and $T_j$ are time-independent coefficients. Plugging Eq. (\ref{sj}) into  Eqs. (\ref{s1}) - (\ref{s4}) leads to,
\begin{equation}\label{a1}
wa_1-(u+p_2w)a_0+ p_1u \sum\limits_{\alpha=1}^{\infty}a_{\alpha N-1}=0,
\end{equation}
\begin{equation}\label{a2}
wa_{(k+1)N}+ua_{kN+(N-2)}+p_2wf(k+1)a_0-(u+w)a_{kN+(N-1)}=0  
\end{equation}
for $k\ge 0$,
\begin{equation}\label{a3}
wa_{kN+1}+(1-p_1)ua_{kN-1}-(u+w)a_{kN}=0  
\end{equation}
for $k \ge 1$ and
\begin{equation}\label{a4}
wa_{kN+i+1}+ua_{kN+i-1}-(u+w)a_{kN+i}=0 
\end{equation}
for $k \ge 0$ and $ i=1,\cdots,N-2$.

Clearly, Eqs. (\ref{a1}) - (\ref{a4}) are identical to the system of equations (\ref{me1}) - (\ref{me4}) for the functions $R_j$  in the $t\rightarrow\infty$ limit, where $dR_j /dt=0$. Thus their solutions should coincide up to the multiplicative constant, namely,
\begin{equation}\label{c}
a_{kN+i}=C R_{kN+i},
\end{equation}
with $R_{kN+i}$ given by Eq. (\ref{rkni}). The normalization condition  $\sum\limits_{k=0}^{\infty}\sum\limits_{i=0}^{N-1} R_{kN+i}=1$ implies that $C=\sum\limits_{k=0}^{\infty}\sum\limits_{i=0}^{N-1} a_{kN+i}$. To find the explicit expression for $C$, we utilize the equations for $T_j$ obtained by plugging Eq. (\ref{sj}) into Eqs. (\ref{s1}) - (\ref{s4}),
\begin{equation}\label{t1}
a_0=wT_1-(u+p_2w)T_0+p_1u\sum\limits_{\alpha=1}^{\infty}T_{\alpha N-1}+p_1u\sum\limits_{\alpha=1}^{\infty}R_{\alpha N-1}-wR_1,
\end{equation}
\begin{eqnarray}\label{t2}
a_{kN+(N-1)}=wT_{(k+1)N}+uT_{kN+(N-2)}+p_2wf(k+1)[T_0-R_0]-(u+w)T_{kN+(N-1)} \nonumber \\ +uR_{kN+(N-2)}-wR_{(k+1)N}
\end{eqnarray}
for $k\ge 0$,
\begin{equation}\label{t3}
a_{kN}=wT_{kN+1}+(1-p_1)uT_{kN-1}-(u+w)T_{kN}+(1-p_1)uR_{kN-1}-wR_{kN+1},
\end{equation}
for $k \ge 1$ and 
\begin{equation}\label{t4}
a_{kN+i}=wT_{kN+i+1}+uT_{kN+i-1}-(u+w)T_{kN+i}+uR_{kN+i-1}-wR_{kN+i+1},
\end{equation} 
for $k \ge 0$ and $ i=1,\cdots,N-2$. Summing up Eqs. (\ref{t1}) - (\ref{t4}) and using $\sum\limits_{k=0}^{\infty}f(k+1)=1$ yields 
\begin{equation}\label{c1}
C=\sum\limits_{k=0}^{\infty}\sum\limits_{i=0}^{N-1} a_{kN+i}=w(1-p_2)R_0+(u-w)=V,
\end{equation}
where the walker's velocity $V$ is given by (\ref{vel1}). Hence, in accordance with Eq. (\ref{c})
\begin{equation}\label{akn}
a_{kN+i}=V R_{kN+i}
\end{equation} 
where $R_{kN+i}$ and $V$ are given by Eqs. (\ref{rkni}) and  (\ref{vel1}).

Now we are able to obtain the expression for the random walker's velocity \cite{ampk}. The mean position of the particle is given by
\begin{eqnarray}\label{avx}
\langle x(t)\rangle=\sum\limits_{m=-\infty}^{+\infty}\sum\limits_{k=0}^{\infty}\sum\limits_{i=0}^{N-1} (mN+kN+i)P_{mN+kN+i,m}(t)= \nonumber\\ \sum\limits_{k=0}^{\infty}\sum\limits_{i=0}^{N-1}\left\{\sum\limits_{m=-\infty}^{+\infty} (mN+kN+i)P_{mN+kN+i,m}(t)\right\}=\sum\limits_{k=0}^{\infty}\sum\limits_{i=0}^{N-1} S_{kN+i}(t),
\end{eqnarray} 
which results in the mean velocity 
\begin{equation}\label{v1}
\tilde{V}=\frac{d}{dt} \langle x(t)\rangle=\sum\limits_{k=0}^{\infty}\sum\limits_{i=0}^{N-1} \frac{d}{dt} S_{kN+i}(t).
\end{equation} 
In the $t\rightarrow\infty$ limit Eqs. (\ref{sj}) and  (\ref{c1}) therefore yield
\begin{equation}\label{vv}
\tilde{V}=\sum\limits_{k=0}^{\infty}\sum\limits_{i=0}^{N-1} a_{kN+i}=w(1-p_2)R_0+(u-w)=V.
\end{equation} 
As expected, Eq. (\ref{vv}) reproduces the expression (\ref{vel1}) for $V$ obtained above with the use of the reduced chemical kinetic scheme method.

For the purposes of computing the diffusion coefficient $D$, functions $T_{kN+i}$ need to be found from (\ref{t1}) - (\ref{t4}) \cite{ampk,derrida83}. In (\ref{t1}) - (\ref{t4}), $a_{kN+i}$ should be expressed according to (\ref{akn}). In analogy with finding $R_{kN+i}$ from (\ref{me1}) - (\ref{me4}), we start with solving (\ref{t4}). In the expression (\ref{rkni}) for $R_{kN+i}$, we denote
\begin{equation}\label{xi}
\xi=\frac{(1-p_2)(1-y)}{1-\beta^N},
\end{equation}  
so that 
\begin{equation}\label{rkni2}
R_{kN+i}=R_{kN}[1-\xi(1-\beta^i)],
\end{equation} 
where $R_{kN}=R_0y^k$, with $R_0$, $y$ given by (\ref{r0}), (\ref{y}). With the use of (\ref{rkni2}), (\ref{t4}) takes the form
\begin{equation}\label{t44}
wT_{kN+i+1}+uT_{kN+i-1}-(u+w)T_{kN+i}+(u-w-V)R_{kN}(1-\xi)+(w-u-V)R_{kN}\xi\beta^i=0.
\end{equation} 
We seek the solution of (\ref{t44}) in the form 
\begin{equation}\label{t45}
T_{kN+i}=C_1(k)+C_2(k)\beta^i+iA(k)+iB(k)\beta^i.
\end{equation} 
In (\ref{t45}), $C_1(k)+C_2(k)\beta^i$ part is the solution of homogeneous equation [the part of (\ref{t44}) which involves only $T_{kN+i}$, $T_{kN+i\pm 1}$], in analogy with Eq. (\ref{c1c2}). Substituting Eq. (\ref{t45}) into Eq. (\ref{t44}) gives these expressions for $A(k)$ and $B(k)$:
\begin{equation}\label{abk}
A(k)=\frac{u-w-V}{u-w}R_{kN}(1-\xi), \ \ \ B(k)=\frac{u-w+V}{u-w}R_{kN}\xi.
\end{equation} 
Plugging $T_{kN}$, $T_{kN+1}$ instead of $T_{kN+i}$ into (\ref{t45}), one can express $C_1(k)$ and $C_2(k)$ in terms of $A(k)$, $B(k)$ as well as $T_{kN}$ and $T_{kN+1}$ (first two points of the period). Substituting resulting expressions for $C_1(k)$ and $C_2(k)$ together with the expressions (\ref{abk}) for $A(k)$, $B(k)$ in (\ref{t45}) gives after some rearrangement
\begin{eqnarray}\label{tkni2}
T_{kN+i}=T_{kN}+\theta_k \frac{1-\beta^i}{1-\beta}+R_{kN}\frac{u-w-V}{u-w}(1-\xi)\left[i-\frac{1-\beta^i}{1-\beta}\right] \nonumber \\ +R_{kN}\frac{u-w+V}{u-w}\xi\left[i\beta^i-\beta\frac{1-\beta^i}{1-\beta}\right],
\end{eqnarray} 
where
\begin{equation}\label{}
\theta_k=T_{kN+1}-T_{kN}.
\end{equation}
Eq. (\ref{tkni2}) solves (\ref{t44}) [i.e., the equation (\ref{t4})] for arbitrary $T_{kN}$ and $\theta_k$. Parameters $T_{kN}$ and $\theta_k$ are to be determined from the Eqs.  (\ref{t2}) and  (\ref{t3}). To get a less cumbersome result for $T_{kN+i}$, it is convenient to rewrite Eq. (\ref{t2}) by adding and subtracting $wT_{kN+N}$, $wR_{kN+N}$ [we note that $T_{kN+N}$, $R_{kN+N}$ are given by (\ref{tkni2}) and (\ref{rkni2}) with $i=N$, i.e., $T_{kN+N}\neq T_{(k+1)N}$, $R_{kN+N}\neq R_{(k+1)N}$]:
\begin{eqnarray}\label{}
a_{kN+(N-1)}=wT_{(k+1)N}-wT_{kN+N}+wT_{kN+N}+uT_{kN+(N-2)}+p_2wf(k+1)[T_0-R_0] \nonumber \\ -(u+w)T_{kN+(N-1)}+uR_{kN+(N-2)}-wR_{kN+N}+w(R_{kN+N}-R_{(k+1)N}),
\end{eqnarray}
which gives 
\begin{eqnarray}\label{t22}
w(T_{(k+1)N}-T_{kN+N})+w(R_{kN+N}-R_{(k+1)N})+p_2wf(k+1)[T_0-R_0]=0,
\end{eqnarray}
as follows from (\ref{t4}) which was formally extended to include $i=N-1$. In (\ref{t22}), the function $f(k+1)$ is expressed according to Eq. (\ref{f(k)2}). Plugging  (\ref{tkni2}), (\ref{rkni2}) (with appropriate $k$, $i$) into (\ref{t22}) permits us to express $\theta_k$ in terms of $T_{kN}$ in this way, 
\begin{equation}\label{theta2}
\theta_k=\frac{1-\beta}{1-\beta^N}\left\{T_{(k+1)N}-T_{kN}-FR_{kN}+p_2y^k(1-y)T_0\right\},
\end{equation}
where 
\begin{equation}\label{f2}
F=\frac{u-w-V}{u-w}(1-\xi)\left[N-\frac{1-\beta^N}{1-\beta}\right]+\frac{u-w+V}{u-w}\xi\left[N\beta^N-\beta\frac{1-\beta^N}{1-\beta}\right].
\end{equation}
It should be mentioned that substituting (\ref{tkni2}) and (\ref{rkni2}) into original Eq. (\ref{t2}) leads to a more involved expression connecting $\theta_k$ and $T_{kN}$ which turns out to be identical to  Eq. (\ref{theta2}), as was numerically verified. 

 To find $T_{kN}$, we substitute (\ref{tkni2}) and (\ref{rkni2}) (with appropriate $k$, $i$) in Eq. (\ref{t3}), with $\theta_k$ in (\ref{tkni2}) replaced according to (\ref{theta2}). After some algebra, this results in 
\begin{equation}\label{abcd1}
aT_{kN}+bT_{(k+1)N}+cT_{(k+2)N}+\{d+ey+fy^2\}T_0y^k+\chi y^k=0,
\end{equation}
where coefficients $a$, $b$, $c$, $d$, $e$, $f$ are given by Eqs. (\ref{a}) - (\ref{f}) and 
\begin{eqnarray}\label{chi}
\chi=-R_0F\left[wy\frac{1-\beta}{1-\beta^N}+(1-p_1)u\frac{1-\beta^{N-1}}{1-\beta^N}\right]+R_0\{(1-p_1)u\phi-wy[1-\xi(1-\beta)] \nonumber \\ +(1-p_1)u[1-\xi(1-\beta^{N-1})]-Vy\}. 
\end{eqnarray} 
In Eq. (\ref{chi}) it was found that 
\begin{equation}\label{phi}
\phi=\frac{u-w-V}{u-w}(1-\xi)\left[N-1-\frac{1-\beta^{N-1}}{1-\beta}\right]+\frac{u-w+V}{u-w}\xi\left[(N-1)\beta^{N-1}-\beta\frac{1-\beta^{N-1}}{1-\beta}\right]
\end{equation}
and $F$ is given by (\ref{f2}).

Comparison between (\ref{abcd1}) and (\ref{abc1}) shows that they are identical (with replacement $R\rightarrow T$), with the exception of $\chi y^k$ term in (\ref{abcd1}). This implies that 
\begin{equation}\label{t(1)}
T^{(1)}_{kN}=\tilde{B}y^k
\end{equation}
with arbitrary constant $\tilde{B}$ solves homogeneous equation [Eq. (\ref{abcd1}) without $\chi y^k$ term], since $y$ given by (\ref{y}) is constructed for $R_{kN}=R_0y^k$ to solve specifically Eq. (\ref{abc1}). 

To account for $\chi y^k$ term, we look for solution of (\ref{abcd1}) of the form 
\begin{equation}\label{t(2)}
T^{(2)}_{kN}=\tilde{C}z^k,
\end{equation}
where 
\begin{equation}\label{z}
z=\frac{-b-\sqrt{b^2-4ac}}{2c}
\end{equation}
with $a$, $b$, $c$ given by (\ref{a}) - (\ref{eq_c}); parameter $\tilde{C}$ is to be determined. Parameter $z$ is the solution of equation $cz^2+bz+a=0$ such that $0\le z<1$. Plugging (\ref{t(2)}) into (\ref{abcd1}) gives
\begin{equation}\label{eqzy}
\tilde{C}z^k(a+bz+cz^2)+(d+ey+fy^2)\tilde{C}y^k+\chi y^k=0.
\end{equation}
The first term in (\ref{eqzy}) vanishes since $z$ is given by (\ref{z}), thus (\ref{eqzy}) yields  
\begin{equation}\label{ctilda}
\tilde{C}=\frac{-\chi}{d+ey+fy^2}.
\end{equation}
The $T^{(2)}_{kN}$ given by (\ref{t(2)}) with $\tilde{C}$ specified by (\ref{ctilda}) therefore solves Eq. (\ref{abcd1}). More generally, (\ref{abcd1}) is solved by the sum of contributions (\ref{t(1)}) and (\ref{t(2)}), i. e. 
\begin{equation}\label{tkn}
T_{kN}=\tilde{B}y^k+\tilde{C}z^k,
\end{equation}
where $\tilde{B}$ remains undetermined. Plugging (\ref{tkn}) in Eq. (\ref{theta2}) for $\theta_k$ gives 
\begin{equation}\label{theta3}
\theta_k=\frac{1-\beta}{1-\beta^N}\left\{\tilde{B}y^k(y-1)+\tilde{C}z^k(z-1)-FR_{0}y^k+p_2y^k(1-y)[\tilde{B}+\tilde{C}]\right\}.
\end{equation}
Substituting (\ref{tkn}) and (\ref{theta3}) into (\ref{tkni2}) results in the final expression for $T_{kN+i}$,
\begin{eqnarray}\label{tkni3}
T_{kN+i}=\tilde{B}y^k+\tilde{C}z^k+\frac{1-\beta^i}{1-\beta^N}\left\{\tilde{B}y^k(y-1)+\tilde{C}z^k(z-1)-FR_{0}y^k+p_2y^k(1-y)[\tilde{B}+\tilde{C}]\right\} \nonumber \\+R_{0}y^k\left\{\frac{u-w-V}{u-w}(1-\xi)\left[i-\frac{1-\beta^i}{1-\beta}\right]+\frac{u-w+V}{u-w}\xi\left[i\beta^i-\beta\frac{1-\beta^i}{1-\beta}\right]\right\}.
\end{eqnarray} 
Although Eq. (\ref{t1}) was not utilized to obtain (\ref{tkni3}), it was
numerically checked that every equation in the system (\ref{t1}) - (\ref{t4}) 
holds for $T_{kN+i}$, $R_{kN+i}$ given by (\ref{tkni3}) and (\ref{rkni}).

To compute the diffusion coefficient $D$, we need to consider additional auxiliary functions \cite{ampk,derrida83},
\begin{equation}\label{ukn}
U_{kN+i}(t)=\sum\limits_{m=-\infty}^{+\infty}(mN+kN+i)^2 P_{mN+kN+i,m}(t), \quad k \ge 0, \quad  i=0,1,\cdots,N-1.
\end{equation} 
A system of equations which determine the time evolution of $U_j(t)$ is derived with the use of Eqs. (\ref{ukn}), (\ref{s0}) and (\ref{r00}) for $U_{kN+i}(t)$, $S_{kN+i}(t)$ and $R_{kN+i}(t)$, as well as Eqs. (\ref{p1}) - (\ref{p4}). It follows that 
\begin{eqnarray}\label{u1}
\frac{d U_{0}(t)}{d t}=wU_1(t)-(u+p_2w)U_0(t)-2wS_1(t)+wR_1(t)+p_1u\sum\limits_{\alpha=1}^{\infty}U_{\alpha N-1}(t) \nonumber \\
+2p_1u\sum\limits_{\alpha=1}^{\infty}S_{\alpha N-1}(t)+p_1u\sum\limits_{\alpha=1}^{\infty}R_{\alpha N-1}(t),
\end{eqnarray}
\begin{eqnarray}\label{u2}
\frac{d U_{kN+(N-1)}(t)}{d t}=wU_{(k+1)N}(t)+uU_{kN+(N-2)}(t)-(u+w)U_{kN+(N-1)}(t)  \nonumber \\ +p_2wf(k+1)U_0(t)-2wS_{(k+1)N}(t)
+2uS_{kN+(N-2)}(t) \nonumber \\ -2p_2wf(k+1)S_0(t)+wR_{(k+1)N}(t) 
+uR_{kN+(N-2)}(t) \nonumber \\ +p_2wf(k+1)R_0(t)
\end{eqnarray}
for $k \ge 0$,
\begin{eqnarray}\label{u3}
\frac{d U_{kN}(t)}{d t}=wU_{kN+1}(t)+(1-p_1)uU_{kN-1}(t)-(u+w)U_{kN}(t)-2wS_{kN+1}(t) \nonumber \\
+2(1-p_1)uS_{kN-1}(t)+wR_{kN+1}(t)+(1-p_1)uR_{kN-1}(t)
\end{eqnarray}
for $k \ge 1$, and 
\begin{eqnarray}\label{u4}
\frac{d U_{kN+i}(t)}{d t}=wU_{kN+i+1}(t)+uU_{kN+i-1}(t)-(u+w)U_{kN+i}(t)-2wS_{kN+i+1}(t) \nonumber \\
+2uS_{kN+i-1}(t)+wR_{kN+i+1}(t)+uR_{kN+i-1}(t)
\end{eqnarray} 
for $k \ge 0$ and $ i=1,\cdots,N-2$.

The diffusion constant is to be found from
\begin{equation}\label{D}
D=\frac{1}{2} \lim\limits_{t\rightarrow\infty}^{} \frac{d}{dt}\left[\langle x(t)^2\rangle -\langle x(t)\rangle^2\right].
\end{equation} 
Using (\ref{ukn}) and summing up Eqs. (\ref{u1}) - (\ref{u4}) results in 
\begin{eqnarray}\label{dx2}
\frac{d}{dt}\langle x(t)^2\rangle=\frac{d}{dt}\sum\limits_{m=-\infty}^{+\infty}\sum\limits_{k=0}^{\infty}\sum\limits_{i=0}^{N-1} (mN+kN+i)^2 P_{mN+kN+i,m}(t)= \nonumber \\ \frac{d}{dt}\sum\limits_{k=0}^{\infty}\sum\limits_{i=0}^{N-1}U_{kN+i}(t)=\sum\limits_{k=0}^{\infty}\sum\limits_{i=0}^{N-1}\frac{dU_{kN+i}(t)}{dt}= \nonumber \\(u+w)-w(1-p_2)R_0(t)+2w(1-p_2)S_0(t)+2(u-w)\sum\limits_{k=0}^{\infty}\sum\limits_{i=0}^{N-1}S_{kN+i}(t),
\end{eqnarray} 
thus we do not need to find $U_{kN+i}(t)$ from the system (\ref{u1}) - (\ref{u4}) to obtain the diffusion coefficient. To derive (\ref{dx2})  we used the normalization conditions, $\sum\limits_{k=0}^{\infty}\sum\limits_{i=0}^{N-1}R_{kN+i}(t)=1$ and $\sum\limits_{k=0}^{\infty}f(k+1)=1$. In the $t\rightarrow\infty$ limit, $S_{kN+i}(t)\rightarrow a_{kN+i}t+T_{kN+i}$, $R_0(t)\rightarrow R_0$, thus (\ref{dx2}) becomes
\begin{equation}\label{dx2dt}
\frac{d}{dt}\langle x(t)^2\rangle=(u+w)-w(1-p_2)R_0+2w(1-p_2)[a_0t+T_0]+2(u-w)\sum\limits_{k=0}^{\infty}\sum\limits_{i=0}^{N-1}[a_{kN+i}t+T_{kN+i}].
\end{equation}
Given that in the stationary-state limit $\frac{d}{dt}\langle x(t)\rangle=\sum\limits_{k=0}^{\infty}\sum\limits_{i=0}^{N-1}a_{kN+i}=w(1-p_2)R_0+(u-w)=V$ [Eq. (\ref{vv})] and utilizing $\langle x(t)\rangle$ given by (\ref{avx}), it follows that 
\begin{eqnarray}\label{dxdt2}
\frac{d}{dt}\langle x(t)\rangle^2 = 2\langle x(t)\rangle\frac{d}{dt}\langle x(t)\rangle = 2V \langle x(t)\rangle=2V \sum\limits_{k=0}^{\infty}\sum\limits_{i=0}^{N-1}S_{kN+i}(t\rightarrow\infty)= \nonumber \\
2[w(1-p_2)R_0+(u-w)] \sum\limits_{k=0}^{\infty}\sum\limits_{i=0}^{N-1}\left\{a_{kN+i}t+T_{kN+i}\right\}.
\end{eqnarray} 
Plugging (\ref{dx2dt}) and (\ref{dxdt2}) into (\ref{D}) gives the diffusion constant,
\begin{equation}\label{DD}
D=\frac{1}{2} \left[(u+w)-w(1-p_2)R_0+2w(1-p_2)[a_0t+T_0]-2w(1-p_2)R_0\sum\limits_{k=0}^{\infty}\sum\limits_{i=0}^{N-1}\{a_{kN+i}t+T_{kN+i}\}\right].
\end{equation} 
The time-dependent part of (\ref{DD}) is
\begin{equation}\label{Dt}
\tilde{D}(t)=\frac{1}{2} \left[2w(1-p_2)a_0t-2w(1-p_2)R_0\sum\limits_{k=0}^{\infty}\sum\limits_{i=0}^{N-1}a_{kN+i}t\right].
\end{equation} 
Using $\sum\limits_{k=0}^{\infty}\sum\limits_{i=0}^{N-1}a_{kN+i}=V$ [Eq. (\ref{vv})] and $a_0=VR_0$ [Eq. (\ref{akn})], it follows that according to (\ref{Dt}) $\tilde{D}(t)=0$. Therefore, as anticipated, the diffusion constant does not contain time-dependent terms, and it is given by
\begin{equation}\label{DD2}
D=\frac{1}{2} \left[(u+w)-w(1-p_2)R_0+2w(1-p_2)T_0-2w(1-p_2)R_0\sum\limits_{k=0}^{\infty}\sum\limits_{i=0}^{N-1}T_{kN+i}\right].
\end{equation} 
In order to get the final expression for diffusion coefficient, it is necessary to calculate $\sum\limits_{k=0}^{\infty}\sum\limits_{i=0}^{N-1}T_{kN+i}$ in (\ref{DD2}). Using Eq. (\ref{tkni3}) for $T_{kN+i}$ yields,
\begin{eqnarray}\label{sumtkni}
\sum\limits_{k=0}^{\infty}\sum\limits_{i=0}^{N-1}T_{kN+i}=\frac{\tilde{B}N}{1-y}+\frac{\tilde{C}N}{1-z}+\left[\frac{N}{1-\beta^N}-\frac{1}{1-\beta}\right]\left\{-\tilde{B}-\tilde{C}-\frac{R_0F}{1-y}+p_2(\tilde{B}+\tilde{C})\right\} \nonumber \\ +\lambda,
\end{eqnarray}  
where 
\begin{eqnarray}\label{lambda}
\lambda=\frac{R_0}{1-y}\left\{\frac{u-w-V}{u-w}(1-\xi)\left[\frac{N(N-1)}{2}-\frac{1}{1-\beta}\left(N-\frac{1-\beta^N}{1-\beta}\right)\right]\right. \nonumber \\ \left.+\frac{u-w+V}{u-w}\xi\left[\frac{-N\beta^N+(N-1)\beta^{(N+1)}+\beta}{(1-\beta)^2}-\frac{\beta}{1-\beta}\left(N-\frac{1-\beta^N}{1-\beta}\right)\right]\right\}.
\end{eqnarray} 
In deriving (\ref{sumtkni}), we used the fact that $0\le y<1$, $0\le z<1$. Next, we consider the last two terms in (\ref{DD2}), 
\begin{eqnarray}\label{2terms}
2w(1-p_2)T_0-2w(1-p_2)R_0\sum\limits_{k=0}^{\infty}\sum\limits_{i=0}^{N-1}T_{kN+i}=2w(1-p_2)\left\{\tilde{B}+\tilde{C}-R_0\left(\frac{\tilde{B}N}{1-y}\right.\right. \nonumber \\ \left.\left.+\frac{\tilde{C}N}{1-z}+\left[\frac{N}{1-\beta^N}-\frac{1}{1-\beta}\right]\left[-\tilde{B}-\tilde{C}-\frac{R_0F}{1-y}+p_2(\tilde{B}+\tilde{C})\right]+\lambda\right)\right\},
\end{eqnarray} 
where we used (\ref{sumtkni}) and $T_0=\tilde{B}+\tilde{C}$ [Eq. (\ref{tkn})]. The contribution from terms $\propto \tilde{B}$ in (\ref{2terms}) can be shown to be
\begin{equation}\label{bterm}
2w(1-p_2)\tilde{B}\left\{1-\frac{R_0}{1-y}\left(N-(1-p_2)(1-y)\left[\frac{N}{1-\beta^N}-\frac{1}{1-\beta}\right]\right)\right\}.
\end{equation} 
Utilizing Eq. (\ref{r0}) for $R_0$, it follows from (\ref{bterm}) that $\tilde{B}$-contribution in (\ref{2terms}) equals $0$, thus undetermined constant $\tilde{B}$ cancels out in (\ref{DD2}) and it has no effect on the diffusion coefficient. Without $\tilde{B}$-terms Eq. (\ref{2terms}) becomes
\begin{eqnarray}\label{2terms2}
2w(1-p_2)\left[T_0-R_0\sum\limits_{k=0}^{\infty}\sum\limits_{i=0}^{N-1}T_{kN+i}\right]=2w(1-p_2)\left\{\tilde{C}-R_0\left(\frac{\tilde{C}N}{1-z}\right.\right. \nonumber \\ \left.\left.-\left[\frac{N}{1-\beta^N}-\frac{1}{1-\beta}\right]\left[\frac{R_0F}{1-y}+(1-p_2)\tilde{C}\right]+\lambda\right)\right\}.
\end{eqnarray} 
Plugging (\ref{2terms2}) into (\ref{DD2}) gives
\begin{eqnarray}\label{DD3}
D=\frac{1}{2} \left\{(u+w)-w(1-p_2)R_0+2w(1-p_2)\tilde{C}-2w(1-p_2)R_0\left(\frac{\tilde{C}N}{1-z}\right.\right. \nonumber \\ \left.\left.-\left[\frac{N}{1-\beta^N}-\frac{1}{1-\beta}\right]\left[\frac{R_0F}{1-y}+(1-p_2)\tilde{C}\right]+\lambda\right)\right\}.
\end{eqnarray} 
In Eq. (\ref{DD3}) we have $\beta=u/w$, and parameters $R_0$, $\tilde{C}$, $y$, $z$, $F$, $\lambda$ are given by Eqs. (\ref{r0}), (\ref{ctilda}), (\ref{y}), (\ref{z}), (\ref{f2}), (\ref{lambda}), correspondingly. Other useful parameters [on which $D$ in (\ref{DD3}) implicitly depends] are $a - f$, $\chi$, $\xi$, $V$ given by Eqs. (\ref{a}) - (\ref{f}), (\ref{chi}), (\ref{xi}), and (\ref{vel1}), correspondingly. With the help of these parameters  Eq. (\ref{DD3}) gives the exact expression for $D$ as a function of transition rates $u$, $w$, probabilities $p_1$, $p_2$, and concentration of bridges $c=1/N$. 

It was verified numerically for various $p_1$ and $c$ values that in the limit of $u\rightarrow 1$, $w\rightarrow 1$, $p_2\rightarrow 0$ Eq. (\ref{DD3}) reproduces the diffusion constant obtained in \cite{ampk} for BBM with $u=w=1$ and $p_2=0$. In the limiting case of $p_1=1$ we obtained $D(u,w,p_2,N)$ in \cite{amk} using Derrida method \cite{derrida83} and it also agrees with our general result (\ref{DD3}) in the $p_1\rightarrow 1$ limit, as was numerically checked.

\section{Discussions}

To illustrate our findings, we plot the dynamic properties of the molecular motor  using Eqs. (\ref{vel1}) and (\ref{DD3}). We first consider the case of the unbiased molecular motor with transition rates $u=w=1$ (Figs. 3 - 5). Eqs. (\ref{vel1}) and (\ref{DD3}) cannot be used directly  when $\beta=u/w =1$. Whereas it is possible to find the $u\rightarrow 1$, $w\rightarrow 1$ limit of Eq. (\ref{vel1}) (although it leads to a cumbersome expression) and thus to plot the velocity, it is problematic in the case of Eq. (\ref{DD3}). Thus we plotted the diffusion
constant for $u=0.999$ and $w=1$ (see  Figs. 3(b), 4(b), 5(b)). Using $u$ values closer to $1$ generates numerical instability. This is a good approximation of the $u=w=1$ case, as was judged from comparison with known limiting cases. Namely, comparing $p_2=0$ case [Figs. 3(b), 5(b)] with the result from \cite{ampk} for $u=w=1$ showed the discrepancy in $D$ values of the order of $ \simeq 0.001$ for almost entire range of parameters $c$ and $p_1$, with the exception of small $c\lesssim 0.02$, and $p_1\lesssim 0.01$, where discrepancy exceeded $0.008$ and $0.006$ correspondingly. For Fig. 4(b), we compared $p_1=1$ case (not shown) with the corresponding case for $u=w=1$ obtained in \cite{amk}: typical discrepancy in $D$ values between $u=0.999$ and $u=1$ cases was $\sim 0.0005$ for all $c$ values except $c\lesssim 0.001$, where discrepancy exceeded $0.007$.

\begin{figure}[tbp]
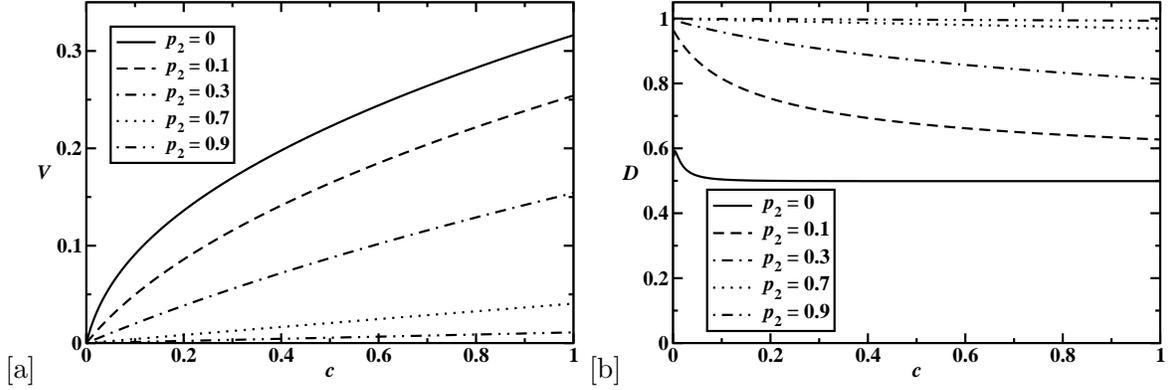
\label{Fig3}
\centering
[a]\includegraphics[scale=0.3,clip=true]{m11vcp2.eps}
[b]\includegraphics[scale=0.3,clip=true]{m11dcp2.eps}
\caption{Dynamic properties of the unbiased molecular motor with $u=w=1$ and with the probability of burning $p_1=0.1$: (a) The mean velocity as a function of the concentration of bridges for different recovery probabilities; (b) the dispersion as a function of the concentration of bridges for different recovery probabilities.   }
\end{figure}

As anticipated, when the recovery probability $p_2\rightarrow 1$ and the presence of bridges has no effect, $V\rightarrow u-w=0$ and  $D\rightarrow\frac{1}{2}(u+w)=1$ for all $c$ (and $p_1$) values (Figs. 3 and 5); the same happens in the limit of the burning probability $p_1\rightarrow 0$ (Fig. 4). Increasing $p_1$ and the concentration of weak links $c$ leads to increasing velocity [as shown in Figs. 3(a), 4(a), 5(a)], whereas increasing $p_2$ reduces the velocity [Figs. 3(a), 5(a)].

\begin{figure}[tbp]
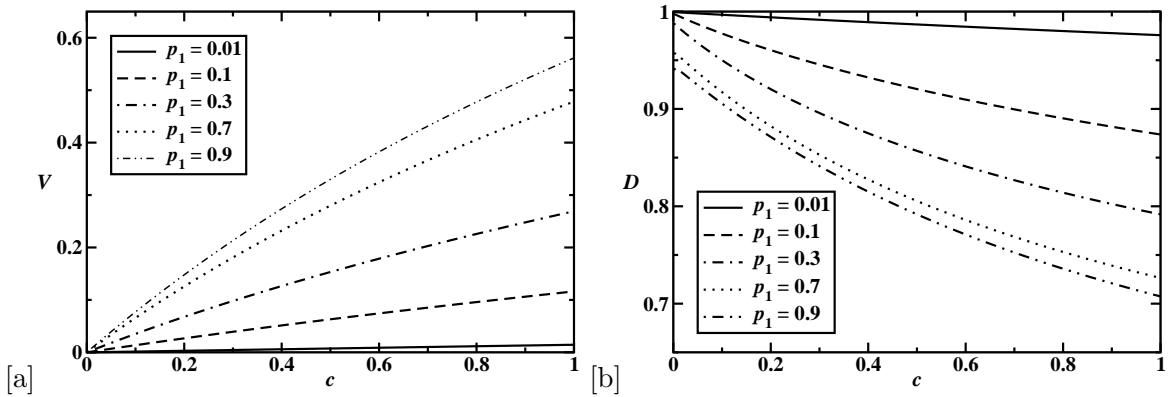
\label{Fig4}
\centering
[a]\includegraphics[scale=0.3,clip=true]{m11vcp1.eps}
[b]\includegraphics[scale=0.3,clip=true]{m11dcp1.eps}
\caption{Dynamic properties of the unbiased molecular motor with $u=w=1$ and with the recovery probability $p_2=0.4$: (a) The mean velocity as a function of the concentration of bridges for different burning probabilities;  (b) the dispersion as a function of the concentration of bridges for different burning probabilities.  }
\end{figure}

\begin{figure}[tbp]\label{Fig5}
\centering
[a]\includegraphics[scale=0.3,clip=true]{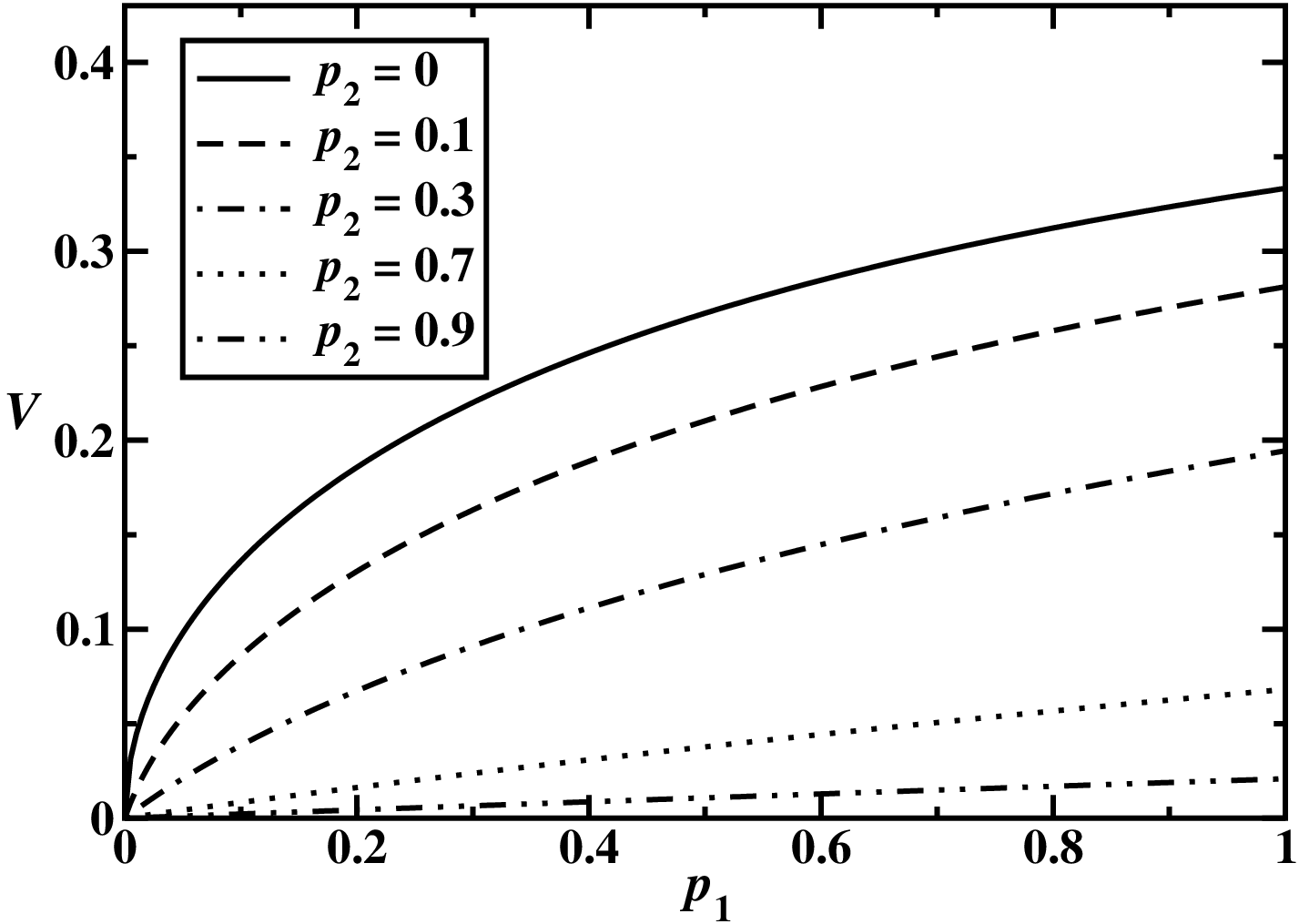}
[b]\includegraphics[scale=0.3,clip=true]{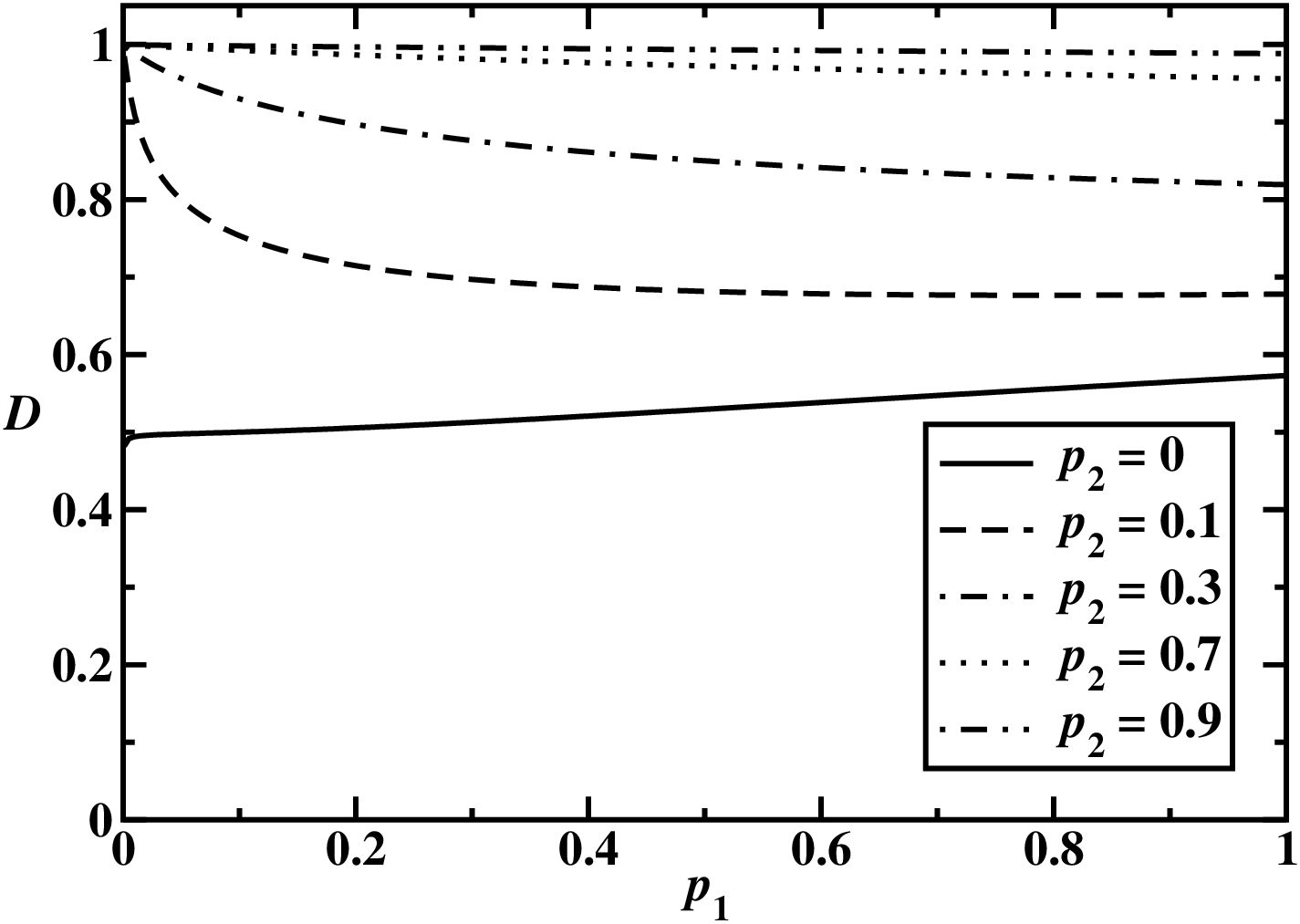}
\caption{Dynamic properties of the unbiased molecular motor with $u=w=1$ and with the concentration of bridges $c=0.2$: (a) The mean velocity as a function of the burning probability for different recovery probabilities;  (b) dispersion as a function of the burning probability for different recovery probabilities. }
\end{figure}

\begin{figure}[tbp]\label{Fig6}
\centering
[a]\includegraphics[scale=0.3,clip=true]{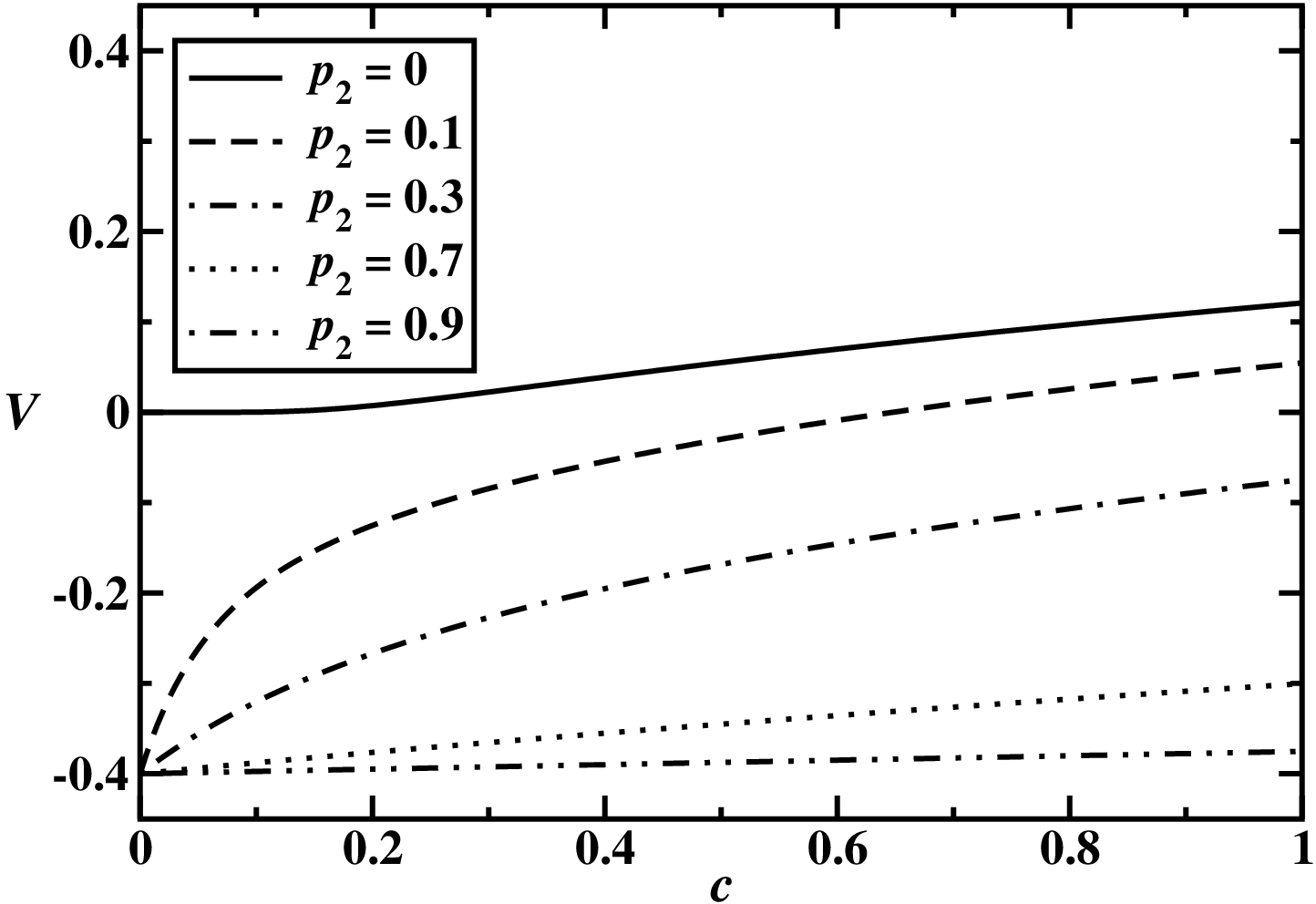}
[b]\includegraphics[scale=0.3,clip=true]{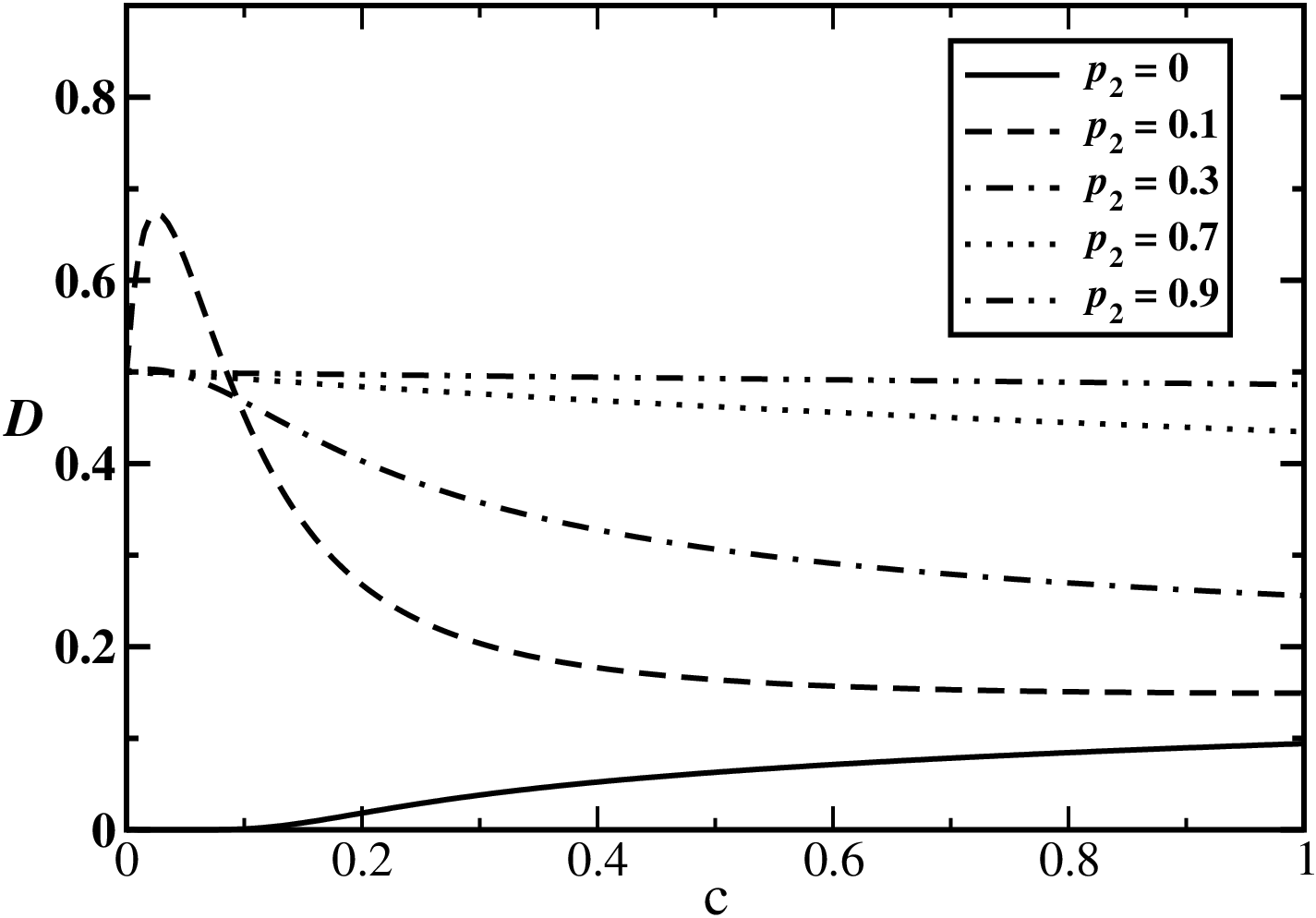}
\caption{Dynamic properties of the backward biased molecular motor with $u=0.3$ and $w=0.7$, and with the burning probability $p_1=0.3$: (a) The mean velocity as a function of the concentration of bridges for different recovery probabilities; (b) the dispersion as a function of the concentration of bridges for different recovery probabilities. }
\end{figure}

\begin{figure}[tbp]\label{Fig7}
\centering
[a]\includegraphics[scale=0.3,clip=true]{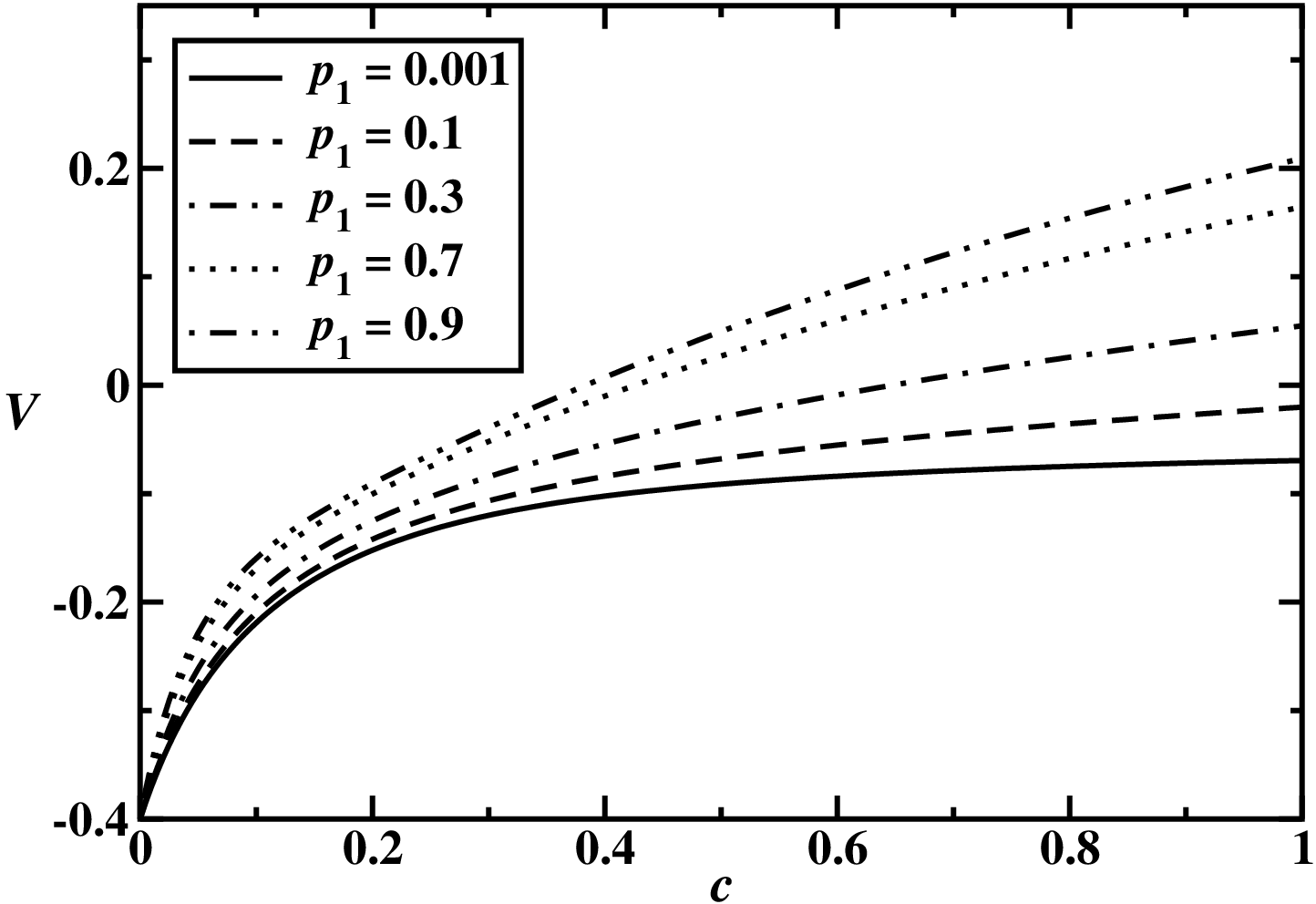}
[b]\includegraphics[scale=0.3,clip=true]{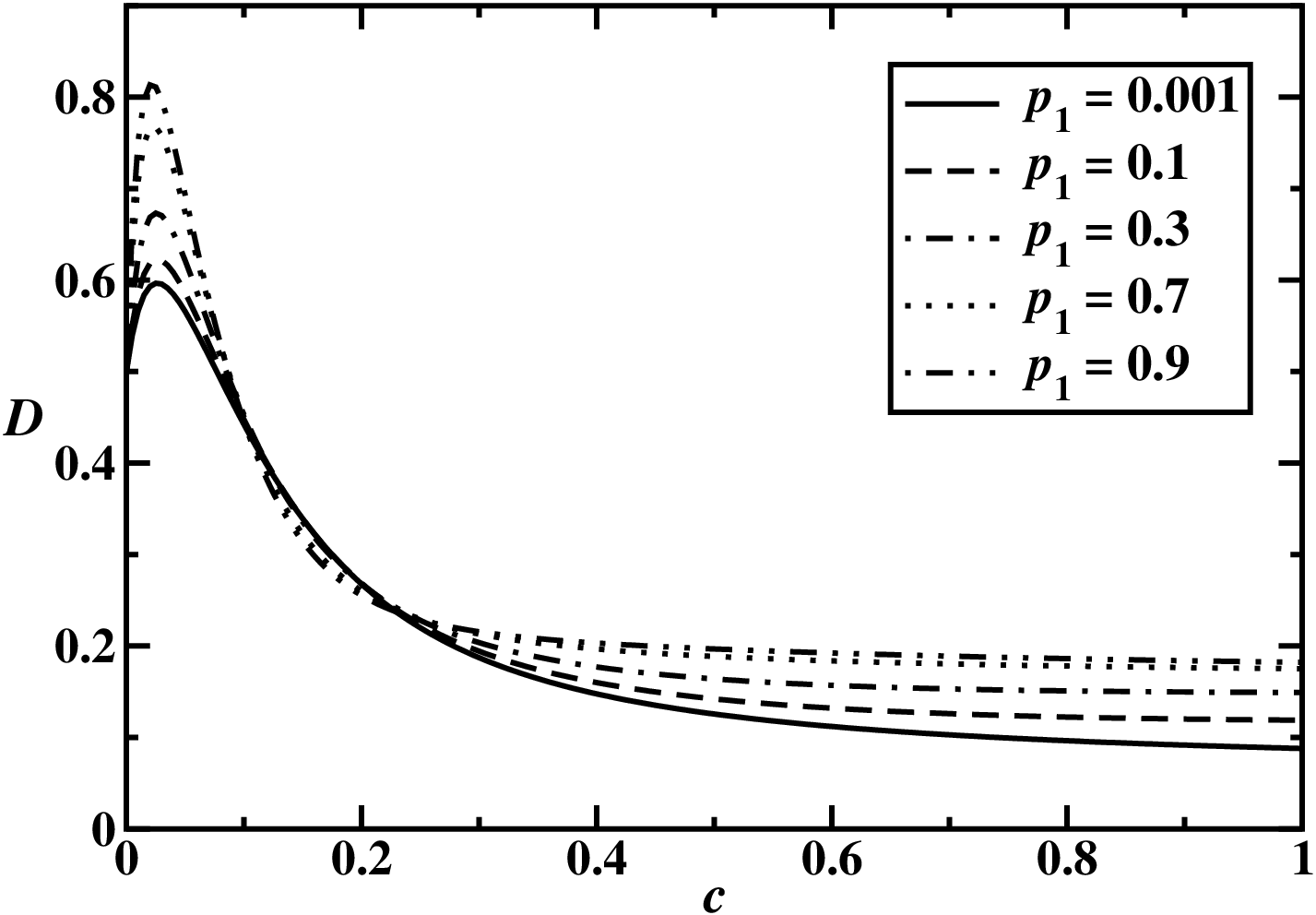}
\caption{Dynamic properties of the backward biased molecular motor with $u=0.3$ and $w=0.7$, and with the recovery probability $p_2=0.1$: (a) The mean velocity as a function of the concentration of bridges for different burning probabilities; (b) the dispersion as a function of the concentration of bridges for different burning probabilities.   }
\end{figure}

The diffusion constant plotted in Figs. 3(b) and 4(b) is a decreasing function of the bridge concentration $c$ because the presence of bridges lowers the fluctuations of the motor protein (the motor protein cannot cross back the already burned bond). We observed that in the limit of low $c$ there is a gap in the dispersion [Figs. 3(b) and 4(b)]: $D(c\rightarrow 0)$ differs from the expected $D(c=0)$ value of $\frac{1}{2}(u+w)=1$. This phenomenon  corresponds to a dynamic transition between unbiased and biased diffusion regimes as was argued earlier in Ref. \cite{amk}. Fig. 3(b) is similar to the corresponding plot in Ref. \cite{amk} for $p_1=1$ case, but for $p_1=0.1$ the gap is prominent only for small $p_2$ values, and  for $p_2>0.1$ it practically disappears. We note that for $p_2=0$ case in Fig. 3(b) the correct $c\rightarrow 0$ limit must be $D(c\rightarrow 0)=2/3$ \cite{ampk}. However, it did not  reach this value because of the emerging numerical instability for very low $c\lesssim 0.001$ (see also our discussion above).

\begin{figure}[tbp]
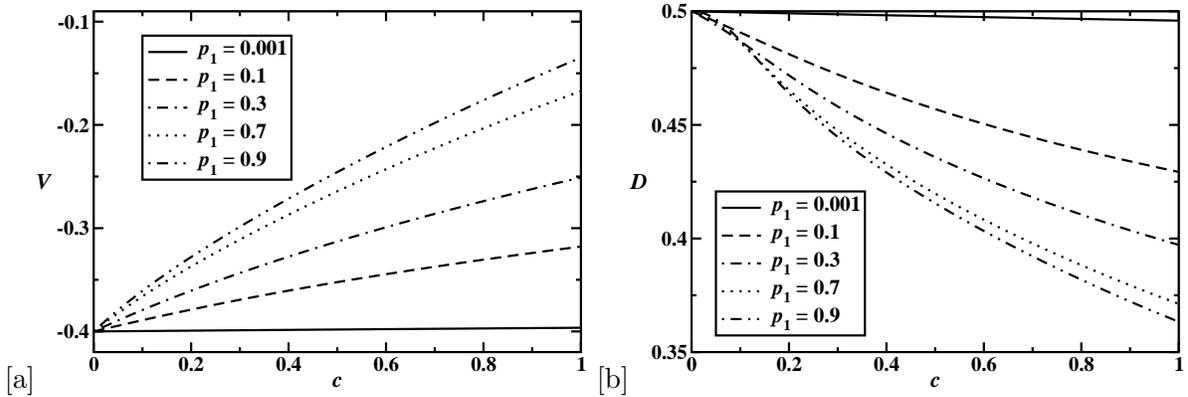
\label{Fig8}
\centering
[a]\includegraphics[scale=0.3,clip=true]{m37vcp1b.eps}
[b]\includegraphics[scale=0.3,clip=true]{m37dcp1b.eps}
\caption{Dynamic properties of the backward biased molecular motor with $u=0.3$ and $w=0.7$, and with the recovery probability $p_2=0.6$: (a) The mean velocity as a function of the concentration of bridges for different burning probabilities;  (b) the dispersion as a function of the concentration of bridges for different burning probabilities. }
\end{figure}

\begin{figure}[tbp]
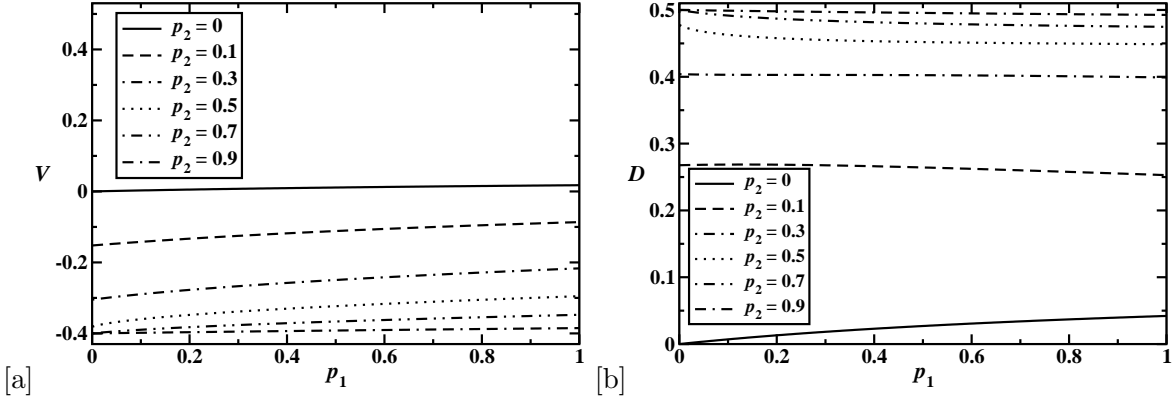
\label{Fig9}
\centering
[a]\includegraphics[scale=0.3,clip=true]{m37vp1p2.eps}
[b]\includegraphics[scale=0.3,clip=true]{m37dp1p2.eps}
\caption{Dynamic properties of the backward biased molecular motor with $u=0.3$ and $w=0.7$, and with the concentration of bridges $c=0.2$: (a) The mean velocity as a function of the burning probability for different recovery probabilities; (b) the dispersion as a function of the burning probability for different recovery probabilities. }
\end{figure}

Analysis of Fig. 5(b) shows that as $p_2$ increases, the behavior of diffusion constant changes from increasing to decreasing function of $p_1$, with $D(p_1)$ developing a minimum for $p_2\lesssim 0.1$. It should be noted that in $p_2=0$ case, $D$ should approach the value of $1/2$ for $p_1\rightarrow 0$ according to \cite{ampk}. We see some discrepancy there which is also due to  the numerical instability for small $p_1$ values. In addition, we observed a gap between $D(p_1\rightarrow 0)=1/2$ for $p_2=0$ and $D(p_1\rightarrow 0)=1$ for all nonzero $p_2$ values.

As another example, we investigated the case of the backward biased motor protein (with specific transition rates $u=0.3, w=0.7$), where bridges are inducing the molecular motor to move in the opposite direction (Figs. 6 - 9). The analysis of dynamic properties shows that in the $p_2\rightarrow 1$ limit (the deterministic bridge recovery) $V\rightarrow u-w=-0.4$ and $D\rightarrow \frac{1}{2}(u+w)=1/2$ for all $c$ and $p_1$ values (see Figs. 6 and 9), as it should be. The $p_1\rightarrow 0$ limit in the $u<w$ case is more complex than in the $u \ge w$ case (when $V\rightarrow u-w$, $D\rightarrow \frac{1}{2}(u+w)$ for all $p_2\neq 0$ and $c$ values as $p_1\rightarrow 0$). Namely, for $u<w$ there exists $\tilde{p}_2(u,w)$ such that for non-zero $p_2<\tilde{p}_2$ the dynamic properties $V(c)$ and $D(c)$ exhibit strong $c$-dependence in the $p_1\rightarrow 0$ limit and they are different from the $u-w$ and $\frac{1}{2}(u+w)$ values for non-zero $c$ [Fig. 7(a),(b)]. Exactly at $p_1=0$ (with $p_2<\tilde{p}_2$), $V(p_1=0,c)=u-w$ and $D(p_1=0,c)=\frac{1}{2}(u+w)$ as obtained using the $y=1$ root of Eq. (\ref{abc2}), given by Eq. (\ref{y}) with ``$+$'' rather than ``$-$'' sign before the square root [the second root of Eq. (\ref{abc2})]. Thus for $p_2<\tilde{p}_2$ there is a dynamic transition  separating $p_1=0$ and $p_1\rightarrow 0$ regimes. For $u=0.3$, $w=0.7$, it was found that $\tilde{p}_2\approx 0.57$. For $p_2>\tilde{p}_2$, $V(p_1\rightarrow 0,c)=V(p_1=0,c)=u-w=-0.4$ and $D(p_1\rightarrow 0,c)=D(p_1=0,c)=\frac{1}{2}(u+w)=0.5$ for all $c$ values [Fig. 8(a),(b)].  Physically this implies that bridges with arbitrarily low burning probability strongly affect the dynamics of the particle which tends to move in the backward direction provided that the recovery probability $p_2$ is less than some critical value; otherwise (in $p_2>\tilde{p}_2$ case) weak links have no effect on the particle dynamics as $p_1\rightarrow 0$. Thus in the $p_1\rightarrow 0$ limit there is a dynamic transition at $p_2=\tilde{p}_2$ separating the regime with $p_2<\tilde{p}_2$ when weak links play a role in the motor protein dynamics and the regime where they are irrelevant ($p_2>\tilde{p}_2$). It should be noted that  the $p_1\rightarrow 0$ limit in Fig. 9 is in agreement with that in Figs. 7 and 8: $\tilde{p}_2\approx 0.57$ plays the same role. The dynamic transition in the $p_1\rightarrow 0$ limit also takes place for $p_2=0$ (irreversible burning of weak connections). It separates backward biased $p_1=0$ regime from forward biased regime with small finite $p_1$, when velocity $V(c)$ is positive for all $c>0$, although $V(c)\rightarrow 0$, $D(c)\rightarrow 0$ for all $c$ values as $p_1\rightarrow 0$ [see Fig. 9(a),(b) with the specific $c$ value]. There are therefore jumps in $V(c)$ and $D(c)$ at $p_1=0$ for all $p_2<\tilde{p}_2$ and all non-zero $c$. 

Increasing the burning probability $p_1$ and concentration of weak bonds $c$ reduces the magnitude of particle's velocity in the backward direction; the same effect is observed when the recovery probability $p_2$ is reduced, as expected [see Figs. 6(a), 7(a), 8(a), 9(a)]. For sufficiently large $c$, $p_1$ (and small $p_2$) the velocity $V$ even becomes positive [Figs. 6(a), 7(a)]. We observed that for $p_2=0$, $V$ is always positive as the burning of weak bonds is irreversible in this case [Figs. 6(a), 9(a)]. In that case, $V(c=0)=u-w$ is different from $V(c\rightarrow 0)=0$ [Fig. 6(a)]. This effect was also observed in Ref. \cite{amk}  in  the $p_1=1$ case.

Diffusion constant demonstrates a more complex behavior, with large fluctuations at small $c$ and small $p_2$ which increase with increasing $p_1$ [Figs. 6(b) and 7(b)]. Fig. 6(b) with $p_1=0.3$ is qualitatively similar to the corresponding figure in \cite{amk} with $p_1=1$, although the maxima in $D$ curves are less pronounced than in the Ref. \cite{amk}. The shape of the $D$ curve differs significantly between Figs. 7(b) and 8(b)  when the threshold $\tilde{p}_2$ is crossed. In case of $p_2=0$ (irreversible bridge burning), fluctuations are reduced (especially at low $c$) and $D$ curve differs substantially from non-zero $p_2$ case [Figs. 6(b) and 9(b)]. As was the case with the velocity, for $p_2=0$ there is a gap in $D$ separating $D(c=0)=\frac{1}{2}(u+w)$ from  $D(c\rightarrow 0)=0$ [Fig. 6(b)], again illustrating a dynamic transition. For non-zero $p_2$  the diffusion constant is $D(c\rightarrow 0)=\frac{1}{2}(u+w)$, and there is no gap [see Figs. 6(b), 7(b), 8(b)].

\section{Conclusions}

We have presented a comprehensive theoretical method of calculating  dynamic properties of molecular motors in reversible burnt-bridge models  for periodic bridge distribution. It is a generalization of the approach developed by us  in  Ref. \cite{mp,ampk} for the unbiased molecular motors and irreversible burning of bridges. Exact and explicit expressions for mean velocity and dispersion have been derived for arbitrary values of parameters $u$, $w$, $p_1$, $p_2$ and $c$. In the known limiting cases of $u=w=1$, $p_2=0$ and of $p_1=1$, we have reproduced our earlier findings \cite{mp,ampk,amk}, thereby confirming the validity of our theoretical analysis. Some interesting phenomena have been observed as a result of the investigation of dynamic properties of the molecular motor in BBM with bridge recovery. It includes dynamic phase transitions and reversal of the direction of the motion. In case of the unbiased molecular motor, increasing the concentration of bridges $c$ (or lowering the recovery probability $p_2$) with other parameters kept fixed results in increasing velocity and decreasing dispersion. However, dependence of the dispersion on burning probability $p_1$ is more complex; it is determined by the $p_2$ value. In the limit of low $c$, gaps in dispersion
plots have been observed for various $p_1$ and $p_2$ values, indicating the dynamic transition between biased and unbiased regimes. Also, a gap was found in the limit of small $p_1$ between $p_2=0$ and non-zero $p_2$ regimes. Thus our results obtained in \cite{amk} for $p_1=1$ with $u=w$ were generalized to cover the full range of $p_1$ values. 

For the  backward biased molecular motor, increasing $c$ has resulted in slowing down the backward movement of the particle, and for sufficiently small $p_2$ (large $p_1$) the direction of motion has been even reversed and the velocity became positive. In the limit of small $p_1$, a dynamic phase transition separating $p_1=0$ and $p_1\rightarrow 0$ regimes has been found provided that $p_2$ is less than some critical value. For sufficiently small $p_2$, broken bridges influence the particle's dynamics even if the burning probability $p_1$ is infinitesimal. The behavior of dispersion as a function of $c$ was non-monotonic for some range of parameters  $p_1$ and $p_2$, with large fluctuations at small $c$ and small $p_2$. In the case of irreversible bridge burning ($p_2=0$), we have observed gaps in velocity and dispersion in $c\rightarrow 0$ limit (for $p_1=0.3$), with the velocity being positive for all non-zero $c$ values. It suggests that there is a dynamic transition at $c=0$ separating backward biased and forward biased diffusion. The velocity and fluctuations are suppressed for sufficiently small $c$. Hence our findings in \cite{amk} for $u<w$ case with $p_1=1$ have been extended to describe the general case of $0<p_1\le 1$. 

The method presented above applies to the case of periodic distribution of weak bonds, which is probably realistic for collagenases \cite{saffarian04}. As a problem to be addressed in the future studies, one can consider BBM with random distribution of bridges \cite{mai01,antal05}  where a different theoretical approach must be applied.


\begin{thebibliography}{99}


\bibitem{AR}  Kolomeisky A B and  Fisher M E 2007 {\it Ann. Rev. Phys. Chem.} {\bf 58} 675 

\bibitem{saffarian04}  Saffarian S,  Collier I E,  Marmer B L, Elson E L and  Goldberg G 2004 {\it Science} {\bf 306} 108  

\bibitem{saffarian06}  Saffarian S,  Qian H,  Collier I E,  Elson E L and  Goldberg G 2006 {\it Phys. Rev. E} {\bf 73} 041909  

\bibitem{mai01}  Mai J,  Sokolov I M and  Blumen A 2001 {\it Phys. Rev. E} {\bf 64} 011102 

\bibitem{antal05}  Antal T and  Krapivsky P L 2005 {\it Phys. Rev. E } {\bf 72} 046104

\bibitem{mp}  Morozov A Y,  Pronina E,  Kolomeisky A B and  Artyomov M N 2007 {\it Phys. Rev. E}  {\bf 75} 031910 

\bibitem{ampk} M.N. Artyomov, A.Y. Morozov, E. Pronina, A.B. Kolomeisky 2007 {\it J. Stat. Mech.} P08002 

\bibitem{mk} A.Y. Morozov and A.B. Kolomeisky 2007 {\it J. Stat. Mech.} P12008

\bibitem{amk} M.N. Artyomov, A.Y. Morozov, A.B. Kolomeisky 2008 {\it
  Phys. Rev. E} {\bf 77} 040901(R)

\bibitem{derrida83}  Derrida B 1983 {\it J. Stat. Phys.} {\bf 31} 433



\end{thebibliography}
\end{document}